\documentclass[twocolumn,showpacs,preprintnumbers,amsmath,amssymb]{revtex4}

\usepackage{amssymb}

\usepackage{graphicx}
\usepackage{dcolumn}
\usepackage{bm}

\begin{document}

\title{Stability of the lattice formed in first-order phase transitions to matter containing strangeness in protoneutron stars.}

\author{J.J. Zach}
 \email{jjzach@pacific.mps.ohio-state.edu}
 \homepage{http://www.physics.ohio-state.edu/~jjzach}
 \affiliation{Department of Physics, The Ohio State University, 174 W. 18th Ave, Columbus, OH, 43210, USA}

\begin{abstract}

Well into the deleptonization phase of a core collapse supernova, 
a first-order phase transition to matter with macroscopic strangeness 
content is assumed to occur and lead to a structured lattice defined 
by negatively charged strange droplets. The lattice is shown to
crystallize for expected droplet charges and separations at
temperatures typically obtained during the protoneutronstar evolution. 
The melting curve of the lattice for small spherical droplets is 
presented. The one-component plasma model proves to be an adequate
description for the lattice in its solid phase with deformation modes 
freezing out around the melting temperature. The mechanical stability 
against shear stresses is such that velocities predicted for convective 
phenomena and differential rotation during the Kelvin-Helmholtz cooling
phase might prevent the crystallization of the phase transition lattice. 
A solid lattice might be fractured by transient convection, which could
result in anisotropic neutrino transport. The melting curve of the 
lattice is relevant for the mechanical evolution of the protoneutronstar 
and therefore should be included in future hydrodynamics simulations.

\end{abstract}

\pacs{26.50.+x,64.60.Cn,97.60.Bw}
\maketitle

\section{Introduction}
\label{intro}

In the past years, considerable effort has gone into the study of the 
existence of phases of strange matter in neutron stars 
\cite{heiselberghjorthjensen,ponsreddy00,ponsmiralles01,balberg99,ghosh95,ponssteiner01}. A macroscopic strangeness content is predicted to be energetically 
favorable at densities well beyond the saturation density of symmetric 
nuclear matter (hereafter $\rho _0$). Conditions with sufficiently high
density and low enough electron chemical potential for the formation of
a macroscopic content of strange quarks (charge $=-1/3$) or particles 
containing such are obtained in the interior of a protoneutronstar 
(hereafter PNS) after post-bounce times of several seconds. 
The three possible forms for the macroscopic manifestation of strangeness 
suggested are a $K^-$ condensate \cite{ponsreddy00,ponsmiralles01}, the 
formation of hyperons \cite{balberg99} and deconfined quark matter 
including strange quarks \cite{ponssteiner01,ghosh95}. The equation of 
state of matter at densities $\rho > \rho_0$ is not well enough known at 
this point to determine which of these scenarios will actually happen 
and what the order of the associated phase transition is. 
\par
The present study limits itself to a certain class of scenarios which
is common to all possible forms of strangeness in high density nuclear 
matter and which might be subject to experimental verification. I assume 
the phase transition to be first order and to result in the formation of 
a lattice in the coexistence region with different energy and charge 
densities of the strange and non-strange phases, as was recently predicted by 
various authors \cite{glendenning92,glendenning99,heiselberg93,glendenning95,glendenning01,christiansen97}. Typical results for the parameters of the mixed phase
lattice, such as the size and spatial separation and the energy and charge 
densities are quoted from these authors. Whereas all studies on this 
lattice to date are limited to cold, deleptonized neutron stars, 
my focus is on PNS's in the Kelvin-Helmholtz cooling phase, 
$\sim 1\,{\rm s} - 30\,{\rm s}$ post-bounce, for temperatures up to
tens of ${\rm MeV}$. In particular, the present paper presents the mechanical 
properties of the structured mixed strange/non-strange phase as they are 
relevant to the further PNS evolution and observational verification.
\par
Many initial studies predicted phenomena caused by strangeness to 
be confined to a very small central region of the PNS 
\cite{baymchin,bethebrowncooperstein}, which would most likely render 
their experimental verification impossible. However, Glendenning showed 
in 1992 \cite{glendenning92} that the previous studies were too simplistic 
in their assumptions about the thermodynamic model of the PNS. In 
particular, there are two conserved charges in PNS matter, the baryon 
number and the electric charge and, therefore, two independent chemical 
potentials, $\mu_e$ and $\mu_B$. During a phase transition, their equality
between the strange (subscript $s$) and non-strange (subscript $n$) phases, 
together with similar conditions for temperature and pressure, comprise 
the Gibbs conditions \cite{glendenning92}: 
\begin{eqnarray}
  \mu _{e,s} = \mu _{e,n} ;&& \mu _{B,s} = \mu _{B,n};\nonumber \\
  T_s=T_n ;&& P_s=P_n.
\end{eqnarray}
Previously, charge neutrality had been enforced in both phases of a 
first-order phase transition separately, and a Maxwell construct had 
been used for the mixed phase: 
$\rho_{mixed} = \chi \rho_s + (1-\chi ) \rho_n$ ($\chi$ being the volume 
fraction of the strange phase), where the densities of each phase 
($\rho_s, \rho_n$) were kept constant throughout the phase transition. 
This additional constraint lead to an overestimate of the required 
density for a phase transition to deconfined (three flavor) quark 
matter of up to $7 \times \rho_0$ as opposed to a predicted range 
of $\sim 2-3 \times \rho_0$ \cite{glendenning92,glendenning99}
with the correct treatment of the Gibbs condition. The microscopic 
stability condition for neutron stars $\delta P / \delta \rho \ge 0$ 
\cite{shapiroteukolsky} and the constant densities in both phases 
also lead to a coexistence region consisting of an infinitely thin spherical 
shell with a pressure discontinuity, as opposed to a spatially 
extended mixed phase with continous $P(r)$-dependence which follows 
from the Gibbs conditions \cite{glendenning95}.
\par
The driving force for a net exchange of charge between the strange and
non-strange phases is the isospin restoring force which leaves the strange 
phase with a net negative charge. Two phases with opposite charges assume a 
spatial order which is determined by a minimal sum of the surface, 
curvature and Coulomb energies \cite{heiselberg93,glendenning95}. The 
predicted geometry of the strange phase has been found 
\cite{glendenning95,glendenning01} to vary from spheres via rods to platelets 
immersed in the majority non-strange phase, as density increases. As soon 
as $\chi >0.5$, the two phases reverse roles. The spatial extent of the 
crystalline phases depends sensitively on the neutron star mass 
\cite{glendenning95} and properties of matter at supernuclear densities 
which are not well known, but can reach several ${\rm km}$. Some of these 
properties have been investigated in recent works, such as the surface 
energy between the strange and hadronic phases and the effective MIT 
bag constant in the case of a phase transition to deconfined quark matter 
\cite{christiansen97} and the surface and curvature energies between normal
nuclear matter and a phase with a $K^-$-condensate \cite{christiansen00}.
In both cases, the results were shown to be model-dependent without a
definitely reliable result. 
\par
The above quoted range of $\sim 2-3 \times \rho_0$ \cite{glendenning92,glendenning99}
for the transition density is based upon the bulk approximation for both 
phases, which neglects any screening effects across the interface. More 
recent studies of surface effects, such as a transition layer with a finite 
thickness of $\Delta R \sim 5\,{\rm fm}$ due to Debye screening effects and 
the discontinous pressure due to the surface tension between both phases,
however, show a net increase in the bulk energy density of the strange phase 
in the case of a $K^-$ condensate \cite{norsen01}, making finite-size droplets for 
radii smaller than $R_S \sim 10\,{\rm fm}$ energetically less favorable. 
Debye screening lengths have been reported for $K^-$ condensed matter 
($\lambda _{K^-} \sim 5\,{\rm fm}$ \cite{norsen01}), for deconfined quark matter 
($\lambda _q \sim 5\,{\rm fm}$ \cite{heiselberg93}) and for hadronic neutron 
star matter ($\lambda _{D,n/p} \sim 10\,{\rm fm}$ and 
$\lambda _{e^-} \sim 13\,{\rm fm}$ \cite{heiselberg93}). Properly taking into 
account screening to find the minimum energy configuration, see \cite{norsen01} 
for $K^-$ condensates and \cite{voskresensky} for deconfined quark matter,
effectively opens up another degree of freedom, in itself lowering the energies 
of both phases, which is favorable for the formation of a lattice. However, 
screening can also increase the electron fraction in the hadronic phase 
by pushing the electrons away from the negative charge on the strange 
droplet surface, leading to a higher negative charge concentration
outside the screened region which will push the pressure at which global 
charge neutrality can be attained to higher values \cite{norsen01}. The 
screened charges have also been shown to increase the effective surface 
tension $\sigma = \sigma_{strong} + \sigma_C $ by a Coulomb contribution, 
making droplets with radii smaller than the Debye screening length energetically 
less favorable (see \cite{voskresensky} for deconfined quark matter), thus
increasing the required PNS density for the first-order phase transition. 
A final answer is not possible unless the exact equation of state for 
hadronic PNS matter with and without a $K^-$ condensate and deconfined 
quark matter with strange quarks is known. This includes hyperon formation, 
following which a similar structured phase has not been studied to date. 
\par
Standard PNS models predict central densities in the relevant range of 
a few times $\rho _0$ after neutrinos have carried away the bulk of the
lepton number in the PNS. A mixed strange/non-strange phase which might form 
then was shown to have a significant impact on the neutrino transport 
properties, resulting in a neutrino opacity up to two orders of magnitude 
higher for typical neutrino energies $\sim 10\,{\rm MeV}$ \cite{reddy00}.
This opens up a window of observation in supernova neutrino detectors through
the remaining lepton number which will be carried away by neutrinos. A pure
strange phase which might form for high enough central densities would
have a lower neutrino opacity compared to non-strange PNS matter
\cite{steiner01}, effectively creating a transparent strange core surrounded 
by a relatively opaque coexistence layer.
\par
In the following, I will refer to a structured mixed layer as phase 
transition lattice, independent upon the specific spatial ordering and 
solid or liquid state. The temperature behavior of the phase transition 
lattice, including the coexistence curve between a solid, crystalline phase 
transition lattice and a liquid phase of the droplets containing strangeness 
is derived in section \ref{temperature}. The mechanical stability of 
the lattice to shear stresses and its possible breakup is investigated 
in section \ref{shearstresses}. Section \ref{signatures} concludes the 
study and gives an outlook on some possible observational signatures of 
the formation of the phase transition lattice. 

\section{The Phase Transition Lattice at Finite Temperatures}
\label{temperature}

\subsection{Phase Transition Lattice Melting Curve}
\label{meltingcurve}

Because a solid, crystalline phase might have to be taken into account
as a new element in hydrodynamical PNS evolution studies and since its 
neutrino transport properties might be different from a liquid mixed phase, 
it is important to know the solid-liquid coexistence curve of the phase 
transition lattice. At the onset of the first-order phase transition and on 
the outermost layers of the mixed phase in the cold neutron star, the 
volume fraction of the strange phase will be small. In that limit, the 
minority phase at zero temperature can be approximated as a Coulomb lattice 
of negative point charges immersed in a slightly positively charged 
background of normal PNS matter. The absolute charge density in the strange 
phase will be higher because the majority phase can significantly lower its 
isospin by pushing negative charge into the minority phase \cite{christiansen97}, 
more than compensating for the opposing effect of a higher repulsive 
Coulomb energy within the latter. The reverse is true for the deepest layers 
of the mixed phase, where the normal PNS matter is the minority phase and 
a high positive charge density resides in that phase, whereas the Coulomb 
interaction is compensated by the condensation energy of hadrons. 
\par
The droplet charges given are to be understood as effective values, already
taking into account screening effects \cite{heiselberg93,norsen01,heiselberg93b}. 
Given rigid strange droplets with charge densities large compared to the surrounding 
hadronic PNS matter, the lattice can then be regarded as a one-component 
plasma (OCP). Its treatment in the harmonic approximation is well-established 
\cite{maradudin71,albers81,chabrier92,chabrier93}. The equation of motion of 
the $\alpha$-component of the displacement $u$ on a lattice site l for a 
Bravais lattice (defined as having one particle per unit cell) can be written as
\begin{equation}
M \ddot{u}_{\alpha }(l) = -\frac{\delta \Phi }{\delta u_{\alpha } (l)} = -\sum _{\beta {l^{\prime}}} \frac{\delta ^2 \Phi}{\delta u_{\alpha }(l) \delta u_{\beta } ({l^{\prime}}) } u_{\beta } ({l^{\prime}} ) ,
\end{equation}
where $\Phi $ is the electrostatic potential and the displacement amplitude 
$u_{\alpha }$ is
\begin{equation}
u_{\alpha }=\sqrt{\frac{\hbar }{2NM}} \sum _{\vec{k} ,j} \frac{e_{\alpha }(\vec{k} j)}{\sqrt{\omega _j(\vec{k})}} e^{i\vec{k}\vec{x}(l)} A_{\vec{k} j},
\end{equation}
with $A_{\vec{k} j} = a^{\dagger}_{-\vec{k} j}+a_{\vec{k} j}$ and the usual 
definitions for creation- and annihilation operators. The three characteristic
polarization modes, two transverse and one longitudinal, of the Bravais lattice 
are denoted by the index $j$. 
\par
In this formalism, the mean square displacement relative to the distance 
between nearest neighbors $d=(3\pi ^2)^{1/6} a$ for a BCC (body-centered cubic) 
lattice is
\begin{eqnarray}
 \frac{\langle u^2\rangle}{d^2} &&= \frac{1}{d^2} \frac{\hbar}{2M} \sum _{\vec{k} j} \frac{coth(\beta \frac{\hbar}{2} \omega_j(\vec{k}))}{\omega_j(\vec{k})} \nonumber\\
&&= \frac{1}{d^2} \frac{3 \hbar}{2 M\alpha \omega _p} (1+\frac{4}{\alpha \eta}D_1(\alpha \eta)),
\end{eqnarray}
where $\eta$ is the degeneracy parameter and the Debye integral is defined as
\begin{equation} 
D_n(x) = \frac{n}{x^n} \int _0 ^x dt(\frac{t^n}{e^t-1}). 
\end{equation}
The dispersion relation for the transverse modes has the acoustic Debye form 
\cite{chabrier92,chabrier93}
\begin{equation}
\omega _T(\vec{k})=\alpha \omega _p \frac{k}{k_D}
\end{equation}
with the Debye wavenumber $k_D=(6\pi ^2 N/V)^{1/3}$, the plasma 
frequency $\omega _p = (Ze/ \epsilon _0 \times N/MV)^{1/2}$ and
$\alpha =0.393$. For the longitudinal branch, the Einstein model 
has been suggested with a constant frequency of 
$\omega _L \propto \omega _p$ \cite{chabrier92}. However,
for a cubic lattice in the harmonic approximation, the symmetry 
condition $\langle u^2 \rangle = 3\times \langle u_T^2 \rangle = 3\times 
\langle u_L^2 \rangle$ \cite{maradudin71} makes only one branch necessary 
to calculate the average square displacement amplitude. For typical 
lattice constants of $a\approx 10\,{\rm fm}$, we obtain 
$k_D \approx 0.4 \,{\rm fm^{-1}}$ and $\hbar \omega _p \approx 5.8\,{\rm MeV}$.
Typical values for the strange (minority) phase were used here, a charge 
density of $\rho _C = 0.4\,{\rm fm^{-3}}$, a mass density of 
$\rho _M = 0.4 \,{\rm fm^{-3}}$ and a droplet radius of $R=3.0\,{\rm fm}$. 
The plasma frequency is an important quantity 
characterizing the lattice, because the degeneracy parameter 
$\eta = \hbar \omega _p/k_B T$ determines the role quantum effects 
play and, ultimately, the freeze-out of the OCP. In the present case, for 
``typical'' protoneutronstar evolution temperatures, $T\sim 10\,{\rm MeV}$, 
we get for the degeneracy parameter 
$\eta = (\hbar Ze)/(\sqrt{M\epsilon _0}k_BTa^{3/2}) \sim 0.5$. 
The problem at hand can therefore be treated neither in the zero 
temperature- (quantum-) nor in the classical limit. 
\par
\begin{figure}[t]
{\centering
{\includegraphics[width=8cm, clip=true, trim=0 0 0 0]{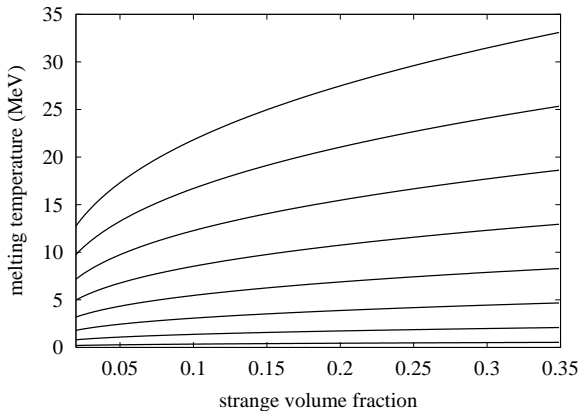}}
\caption{Melting curve for different droplet charge densities: $\rho _C = 0.1 \,{\rm fm^{-3}}$ (bottom curve) to $\rho _C = 0.8 \,{\rm fm^{-3}}$ (top curve) with
$\rho = 0.4\,{\rm fm^{-3}}$ and $R=3.0\,{\rm fm}$. \label{meltingcurve1}}
}
\end{figure}
\begin{figure}[t]
{\centering
{\includegraphics[width=8cm, clip=true, trim=0 0 0 0]{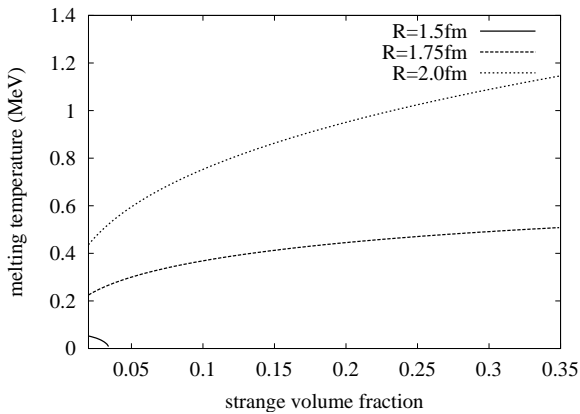}}
\caption{Melting curve for different droplet radii with $\rho _C = 0.4 \,{\rm fm^{-3}}$ and $\rho = 3.0\,{\rm fm^{-3}}$. 
\label{meltingcurve2}}
}
\end{figure}

The Lindemann parameter $\gamma^2$ is defined as the value of the 
quantity $\langle u^2\rangle/d^2$ at the solid-liquid transition.
For an OCP, it has been determined using Monte Carlo - simulations in both 
the classical (high temperature) limit \cite{pollock73,ogata87,stringfellow90} 
and in the quantum case (zero temperature) \cite{ceperley80}, for both 
fermions and bosons. Energetically, spin pairing effects on the droplets
will drive their total spin to zero. I therefore treat them as bosons. For 
the intermediate degeneracies given here, the interpolation formula for 
the Lindemann parameter of a bosonic OCP by Chabrier \cite{chabrier93} 
is used:
\begin{equation}
\gamma (\eta) = \gamma _0 - \frac{0.096+4.31\times 10^{-3} \eta ^2 }{1+0.05 \eta ^2 + 2.092 \times 10^{-4} \eta ^4},
\end{equation} 
with $\gamma _0 = 0.249$ being the quantum limit. The melting curves for 
different charge densities on the strange droplets are plotted in figure 
\ref{meltingcurve1}, where the curves represent charge densities 
from $\rho _C = 0.1\,{\rm fm^{-3}}$ to $\rho _C = 0.8 \,{\rm fm^{-3}}$ in steps 
of $\rho _C = 0.1 \,{\rm fm^{-3}}$ for a mass density of 
$\rho _M = 0.4 \,{\rm fm^{-3}}$ and droplet radii $R=3.0\,{\rm fm}$. If
the droplets are treated as fermions, the result only differs significantly 
for coexistence curves with transition temperatures below $1\,{\rm MeV}$. 
It can be seen that an initially liquid phase transition lattice 
crystallizes for PNS temperatures of $T \sim 1-10\,{\rm MeV}$, which 
lies well within the range obtained during the Kelvin-Helmholtz cooling 
timescales in the standard PNS paradigm, for a wide range of parameters 
($\chi$ and $\rho _C$). There is, for any given charge density, a lower 
limit on the droplet radius below which there is no crystallization. For 
example for $\rho _C = 0.4 \,{\rm fm^{-3}}$, this limit is between 
$1.5-1.75\,{\rm fm}$, see figure \ref{meltingcurve2}. Hence, lattices 
with larger droplet radii solidify earlier in the PNS evolution, causing 
the mixed layer to freeze out from its interior outwards.  

\subsection{Deformation Modes}
\label{deformations}

The OCP assumes negative, inherently rigid, point charges in a sea of 
positive background. However, since the surface tension is only 
$\sigma \sim 10\,{\rm MeV}/{\rm fm^2}$, small compared to strong interaction
energy scales $\epsilon _{strong} \approx 10^3\,{\rm MeV}/{\rm fm^3}$ for
the given droplet dimensions, it is clear that the strange phase droplets 
cannot necessarily be considered as rigid. It is therefore important to 
know whether deformation modes have to be taken into consideration in the 
treatment of lattice vibrations. 
\par
Consider a droplet of strange matter which is slightly elongated along the 
x-direction to $R+dR$, yielding an ellipsoid:
\begin{equation}
 (a,b,b) \sim (R+dR,R/\sqrt{1+\frac{dR}{R}},R/\sqrt{1+\frac{dR}{R}}).
\end{equation} 
Its surface area is
\begin{equation}
 S=2\pi (b^2+\frac{a^2b}{\sqrt{a^2-b^2}} arcsin(\frac{a^2-b^2}{a^2})),
\end{equation}
which, when expanded to second order in $dR$, yields
\begin{equation}
 S\simeq 4\pi R^2 + 2\pi \frac{9}{8} dR^2 = S_0 + \Delta S,
\end{equation}
from which the elastic constant $k_S$ for the deformation energy can be deduced:
\begin{equation}
 \Delta E_S = \sigma \Delta S = 2\pi \frac{9}{8} \sigma dR^2 = \frac{1}{2} k_S dR^2.
\end{equation}
The inertial term $m_S$ can be found via the kinetic energy
\begin{eqnarray}
 \int _{-R} ^{R} dx \int _0 ^{\sqrt{R^2-x^2}} d\rho (2\pi \rho [\frac{1}{2}\rho_M x^2 (\frac{\omega _S}{2\pi })^2]) \nonumber\\
= \frac{1}{30\pi } \rho _M \omega_S^2R^5 = \frac{1}{2} (\frac{4\pi }{15} \rho _M R^3) \dot{R}^2 = \frac{1}{2} m_S \dot{R}^2.
\end{eqnarray}
The characteristic vibration energy can therefore be estimated as
\begin{equation}
 \omega _S = \sqrt{\frac{k_S}{m_S}} = \sqrt{\frac{9\pi \sigma /2}{M/5}}.
\end{equation}
A typical value, for $\rho_M = 0.4\,{\rm fm^{-3}}$, $R=2\,{\rm fm}$ and 
$\sigma = 10\,{\rm MeV}{\rm fm^{-2}}$, is $\hbar \omega _S \sim 10\,{\rm MeV}$, 
which is comparable to the plasma frequency. {\em Hence, deformation modes 
freeze out at about the same temperature as lattice vibrations.} 
The OCP can therefore be considered as a valid description of the melting 
curve of the phase transition lattice, since no other modes are relevant 
once it becomes a crystal. At temperatures above the transition between a 
liquid and a solid strange droplet phase, the lattice- and deformation 
modes will be in thermal equilibrium.

\section{Shear Stresses and the Strange Phase Transition Lattice}
\label{shearstresses}

Besides thermodynamic criteria for the existence of a crystalline mixed phase, 
we need to know whether the lattice can withstand typical shear stresses 
present in convective PNS cores. The shear constant of a cubic Coulomb lattice 
is \cite{fuchs35}
\begin{equation}
  c_{44}=\frac{d^2W_l}{d\gamma ^2_{xy}},
\end{equation}
where $W_l$ is the lattice energy and $\gamma _{xy}$ the angle of distortion. 
The Coulomb lattice energy can be calculated using Ewald's method 
\cite{ewald21,dove}:
\begin{widetext}
\begin{equation}
  W_l = \frac{1}{2} \sum _l \frac{e^2}{4\pi \epsilon _0 r(l)} = \frac{1}{2} \frac{e^2}{4\pi \epsilon _0} \left(\sum _l \frac{\mbox{erfc}(gr(l))}{r(l)} + \sum _l \frac{4\pi }{\Omega} \frac{exp(-G_l^2/4g^2)}{G_l^2} \right), 
\end{equation}
\end{widetext}
where the complementary error function 
$\mbox{erfc}(x)\equiv \frac{2}{\sqrt{\pi }}\int_x^{\infty }exp(-y^2)dy$, 
$\Omega $ is a unit cell volume and $g$ is a parameter to be adjusted for 
fast numerical convergence of both real (vectors $\vec{r}(l)$) and inverse 
(vectors $\vec{G}(l)$) lattice sums. The result for a bcc-lattice is 
\cite{fuchs35} 
\begin{eqnarray}
  c_{44}&&=0.7423\times \frac{\frac{4}{3}\pi R^3 \rho_C e^2}{4\pi \epsilon _0 a}\nonumber\\ && = 4.477\,{\rm MeV}\times (R^3 \rho_C) (\frac{a}{1.0\,{\rm fm}})^{-1},
\end{eqnarray}
$a$ being the lattice constant. The critical shear stress is the force per 
unit area necessary to maintain two planes of the crystal distorted against 
each other by an angle corresponding to a displacement of ${a/4}$ perpendicular
to a lattice plane \cite{ashcroft631}, which is in the linear approximation 
\begin{eqnarray}
  \sigma _{crit} &&=\frac{1}{NA} \frac{dU}{dx} \approx \frac{1}{A} \frac{d}{dx} (2 c_{44} (\frac{x}{a})^2) \nonumber\\&&=\frac{c_{44}}{a^3} = 4.477\,{\rm MeV}\times (R^3 \rho_C) (\frac{a}{1.0\,{\rm fm}})^{-4},
\end{eqnarray}
which, for a set of typical values, a lattice constant $a=10\,{\rm fm}$, droplet
radius $R=2.0\,{\rm fm}$ and charge density $\rho _C=0.4 \,{\rm fm^{-3}}$, gives 
$\sigma _{crit} = 1.4 \times 10^{-3} \,{\rm MeV}{\rm fm^{-3}}$. Stresses in 
that order of magnitude might be caused by convection or differential rotation 
of the newly formed PNS. These stresses can either prevent the formation of
the solid lattice in the first place or, if they are due to phenomena which
are prone to variations, such as convection, break up a solid lattice formed 
during a transient period of weak convection.  
\par
A negative gradient in the lepton concentration has been shown to lead
to convection during the Kelvin-Helmholtz cooling phase of the PNS 
\cite{epstein79}. More recently, hydrodynamics simulations including
convection indicate that the Ledoux criterion for convective instability
\begin{equation}
C_L \equiv \left(\frac{\delta \rho }{\delta S}\right)_{P,Y_l}\frac{dS}{dr} + \left(\frac{\delta \rho }{\delta Y_l}\right)_{P,S}\frac{dY_l}{dr} > 0
\end{equation}
is true in most of the PNS for times of more than $\sim 1\,{\rm s}$ after bounce 
\cite{keil96}. The pure strange phase in the center of the PNS, if it exists, 
is not expected to show strong negative lepton or entropy gradients. This is due 
to the relatively opaque mixed phase enclosing it and the fact that the 
transport of both heat and lepton number in the PNS interior is 
dominated by neutrinos. The most violent convection will therefore take place
in the matter exterior to the phase transition lattice. Although various 
authors disagree on the extent and strength of convection in PNS's, 
convective velocities of $v_c \sim 10^6 \,{\rm m}{\rm s^{-1}}$ are reported 
in many studies \cite{keil96,burrows87,herant94}. This is equivalent to a 
kinetic energy density of $E_{conv}/V \sim 10^{-3} \,{\rm MeV}/{\rm fm^3}$, 
which indicates that convection might indeed be able to either break an
existing phase transition lattice or prevent its formation. A strong enough
convective cell forming outside the mixed strange/non-strange layer after 
its crystallization during a transient quiet period might fracture 
the solid lattice and mix matter from the non-strange envelope into the 
now liquid phase transition lattice in the region below
the convective cell. For the duration of the convective flow, this 
would result in a localized hole with a substantially lowered neutrino 
opacity \cite{reddy00} compared to the still intact solid lattice in 
all other directions, hence in anisotropic neutrino transport through 
the mixed strange/non-strange layer. 
\par
The discovery of a number of millisecond pulsars in recent years 
\cite{cordes97} indicates that some of the angular momentum residing in the
core collapse supernova might remain in the PNS. The resulting rotation
is likely to be differential and has been studied by Goussard et al. 
\cite{goussard97,goussard98} who solved the relativistic stellar structure
equations for rotating PNS's with representative equations of state
for the different epochs in the PNS evolution. The rotation period $\Omega$ 
as a function of radius $r$ assumed in that study is (in the Newtonian limit)
\cite{goussard98}
\begin{equation}
  \Omega = \frac{R_0^2 \Omega _C}{R_0^2+r^2\,sin^2(\theta )},
\end{equation}
with $R_0 \sim 1\,{\rm km}$ the characteristic scale of variation of 
$\Omega $, $r\, sin(\theta )$ the distance from the rotation axis and 
$\Omega _C$ the central rotation period. The rotation period a PNS can 
acquire without additional accretion has been shown to be limited by the 
increase of the minimum neutron star mass for times up to $\sim 100\,{\rm ms}$ 
post-bounce and by the mass shedding limit beyond that, resulting in 
$P_{min} \approx 1.7\,{\rm ms}$ \cite{goussard98}. This corresponds to 
velocities of 
$v \sim (r/1\,{\rm km}) (\Omega / \Omega _C) 10^6\,{\rm m}{\rm s^{-1}}$,
which is comparable in magnitude to convective velocities sufficient to 
cause critical shear stresses. It is not likely that regions of a fast 
rotating PNS at several seconds post-bounce go through a transient phase 
with low rotation period during which the phase transition lattice could 
crystallize, possibly to be broken up at later times by a larger gradient 
in the rotation period. Rather, for PNS's with rotation periods below 
$\sim 100\,{\rm ms}$, the timescale for the stratification of differential 
rotation might be of significant importance for the melting curve of the 
phase transition lattice, possibly comparable to the cooling timescale. 
However, unless the transport of angular momentum in core collapse supernovae
and in particular within the PNS is finally resolved, the melting curves
presented in sec. \ref{meltingcurve} only apply to PNS's with low differential
rotation.

\section{Summary and Discussion}
\label{signatures}

The present study shows that if the phase transition from hadronic 
PNS matter to strange matter is of first order and if it results in a lattice 
of separate strange and non-strange phases in the coexistence zone, 
it will crystallize for temperatures predicted during the Kelvin-Helmholtz 
cooling phase of a PNS following a core collapse supernova. The process of
crystallization has, however, complex interactions with the hydrodynamical
evolution of the PNS. Shear stresses due to strong differential rotation 
with minimum periods in the range observed for millisecond pulsars and 
convection induced by lepton gradients might, as long as they persist, 
prevent the formation of a solid lattice. A final answer can only be given
by the full inclusion of a possibly solid mixed strange/non-strange phase in
future core collapse supernova simulations.
\par
If the first-order phase transition occurs before the end of the 
Kelvin-Helmholtz cooling phase, the remaining neutrinos to be emitted 
might serve as a possible window on its formation and crystallization. 
With the predicted increase in the neutrino opacity for a first 
order phase transition lattice \cite{reddy00}, expected to be especially
pronounced for intermediate neutrino energies in the range 
$\sim 10-100 \,{\rm MeV}$, the timescale for the decay of the neutrino
emission will increase, which might be observable as a knee in the 
neutrino luminosity. Once the lattice becomes solid, the evolution might 
be accompanied by fractures and rearrangements in the mixed strange/non-strange 
zone, possibly showing irregularities in the neutrino luminosity. If 
a changing lepton number gradient causes a convective cell to form in 
a formerly non-convective region located outside a previously crystallized 
phase transition lattice, sufficiently violent convection might break the
solid lattice and locally cause it to return to a liquid state. Non-strange PNS 
matter transported from regions outside the phase transition lattice 
might mix with the liquid, resulting in an anisotropic neutrino-opaque 
layer with a partial hole and therefore anisotropic neutrino transport. 
This will be more pronounced if the central density of the PNS is high 
enough to allow for a pure strange core enclosed by the relatively opaque 
mixed zone which essentially dams up neutrinos behind it. The extent to 
which the phase transition lattice will affect the neutrino luminosity 
and emission spectrum will be the topic of a detailed transport study 
with a special focus on anisotropic neutrino transport in conjunction 
with a hydrodynamical treatment of the non-strange layers above the 
coexistence zone \cite{zach}. This will also include the different neutrino 
transport properties of a solid versus a liquid lattice.

\section{Acknowledgements}

J.J. Zach would like to thank George M. Fuller at the University of 
California at San Diego (UCSD) for the hospitality and valuable 
discussions on the subject during two visits in the spring and fall of 2001. 
These were supported by NSF grant number PHY 9800980 and its renewal. 
Further, many thanks to Richard N. Boyd at The Ohio State University 
(OSU) for fruitful discussions and financial support of this work 
through NSF grant number PHY 9901241. For valuable inputs, I would 
also like to thank Neal Dalal and Kevork Abazajian from UCSD, 
Alexander StJ. Murphy from the University of Edinburgh and Richard 
Furnstahl from OSU. The recommendations by H.-Thomas Janka from the
Max Planck Institute in Garching/Germany concerning possible 
observational consequences of the present work were outstanding and very 
helpful.

\end{document}